\def\kmss    {km\, s$^{-1}$}
\def\kms     {km\, s$^{-1}$ }

\documentclass{iau} 
\usepackage{graphicx}

\title[H$_2$O MegaMasers] 
{H$_2$O MegaMasers:  RadioAstron success story}

\author[Baan, Alakoz, et al.]   
{Willem Baan$^{1,2}$,
Alexey Alakoz$^3$, \\
 Tao An$^4$, Simon Ellingsen$^5$, Christian Henkel$^{6,7}$, Hiroshi Imai$^8$, Vladimir Kostenko$^3$, 
 Ivan Litovchenko$^3$, James Moran$^9$, Andrej Sobolev$^{10}$, and Alexander Tolmachev$^3$
}

\affiliation{$^1$Netherlands Institute for Radio Astronomy, Dwingeloo, The Netherlands, m: {\tt baan@astron.nl} \\
$^2$ XinJiang Astronomical Observatory, Chinese Academy of Sciences, Urumqi, PR China \\
$^3$AstroSpace Center, Lebedev Physical Institute,  Moscow, Russia, m: {\tt alexey.alakoz@gmail.com} \\
$^4$ Shanghai Astrophysical Observatory, Chinese Academy of Sciences, Shanghai, PR China \\
$^5$ University of Tasmania, Hobart, Australia \\
$^6$ Max Planck Institut f\"ur Radioastronomie, Bonn, Germany \\
$^7$ Astron. Dept., King Abdulaziz Univ., Jeddah, Saudi Arabia \\
$^8$ Kagoshima University, Kagoshima, Japan \\
$^9$ Center for Astrophysics, Cambridge MA, USA \\
$^{10}$ Ural Federal University, Ekaterinburg, Russia \\
}

\pubyear{2017}
\volume{336}  
\jname{Astrophysical Masers: Unlocking the Mysteries of the Universe}
\editors{A. Tarchi, M.J. Reid \& P. Castangia, eds.}
\begin{document}

\maketitle

\begin{abstract}
The RadioAstron space-VLBI mission has successfully detected extragalactic H$_2$O MegaMaser 
emission regions at very long Earth to space baselines ranging between 1.4 and 26.7 Earth Diameters (ED).  
The preliminary results for two galaxies, NGC\,3079 and NGC\,4258, at baselines longer than one ED
indicate masering environments and excitation conditions in these galaxies that are distinctly different. 
Further observations of NGC\,4258 at longer baselines will reveal more of the physics 
of individual emission regions.

\end{abstract}

\firstsection 
\section{Introduction}

The Space Radio Telescope or the RadioAstron Observatory (RAO) is an international space-VLBI project led by the 
Astro Space Center of PN Lebedev Physical Institute.
The RadioAstron payload on board of the Spectr-R mission has been equipped  with a 10 meter antenna, two hydrogen masers, and receivers in P, L, C, and K-band \cite{Kardashev2013}.
This paper presents some recent results of RAO observations of H$_2$O MegaMasers (MM) using 
the highest window of the Multi-Frequency Synthesis (MFS) system at 22 GHz, 
which covers a redshift range z = 0.0 - 0.0053 for H$_2$O MM emission studies.
 A total of 24 known H$_2$O MM have a redshift falling within this window but only 7 of them are deemed strong enough for detection with RadioAstron.  
The strong nearby sources that may be searched and their observing status are: 
\\ \\
NGC3079      - 3.5 Jy  - masering material shocked ISM in nucleus - detected at 1.6 - 2.3 ED \\
NGC4258     - 9.8 Jy  - maser regions in compact nuclear disk - detected at 1.4 - 26.7 ED \\
NGC4945      - 8.5 Jy - masers in nuclear disk  - not yet searched \\
N133 and 30Dor  - 70 \& 3 Jy  - star formation regions in LMC - not yet detected \\
Circinus        - 4.2 Jy  - masers in Keplerian disk \& bi-conical outflow - not yet detected \\
IC133      - 1.5 Jy - star formation region in M33 - not yet searched \\
NGC1068    - 0.65 Jy   - Keplerian nuclear disk - potential candidate\\

The initial detections of both NGC\,3079 and NGC\,4258 were obtained from observations in late 2014. 
After the update of the orbital elements for the ASC correlator \cite{LikhachevEA2017},
ten more new detections have been obtained for NGC\,4258 up to a baseline of 26.7 Earth 
Diameters (ED). 

\begin{figure}[h]
\begin{center}
\includegraphics[width=6cm]{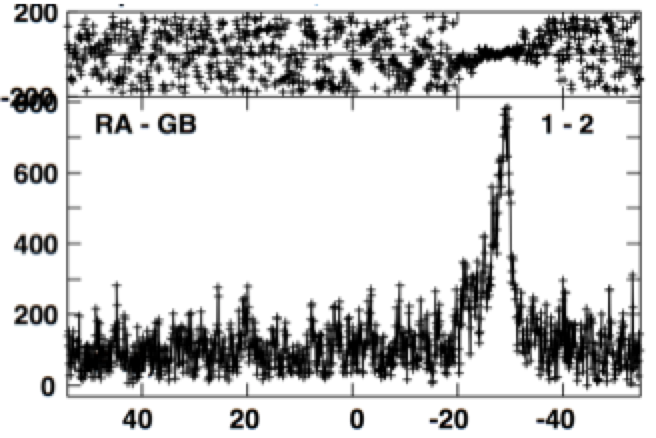}
\includegraphics[width=6cm]{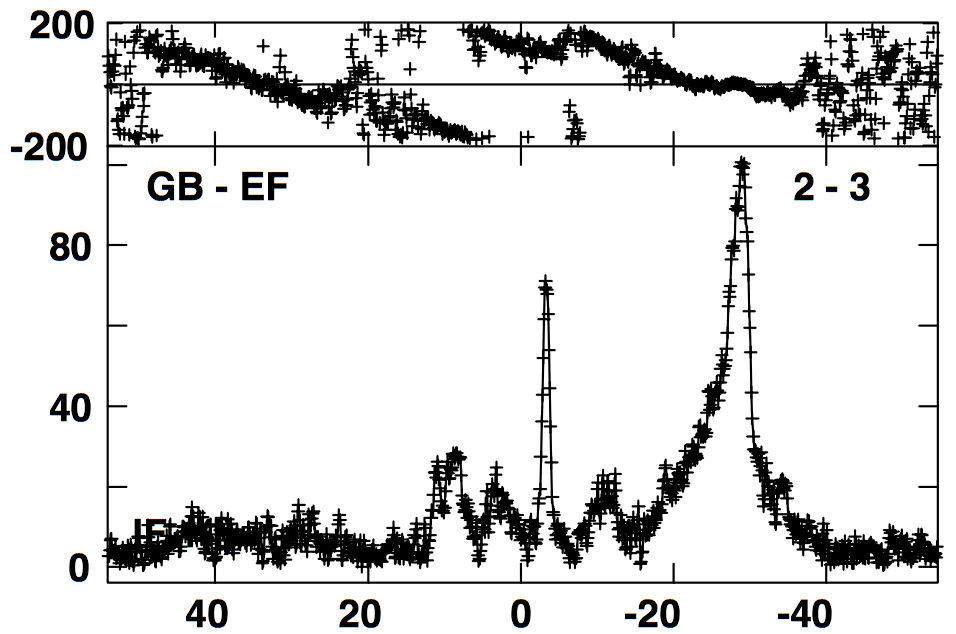} 
 \caption{RadioAstron detection of NGC\,3079 with Green Bank Telescope (GBT) at 2.3 ED. 
 (left) The uncalibrated cross-correlation and fringe phase spectrum on the RAO-GBT baseline. 
 (right) The uncalibrated cross spectrum and the fringe phase on the GBT - Effelsberg baseline. 
 The axes are in arbitrary units of flux density and the velocity scale is centered at 984 \kmss.}
   \label{fig1}
\end{center}
\end{figure}

\section{NGC\,3079 results}

The high-brightness maser components in the H$_2$O MM NGC\,3079 form an arc that is offset 
from the triple components of the Compact Symmetric Object (CSO) at the nuclear center \cite{KondratkoEA2005}.  
The systemic velocity of NGC\,3079 is 1116 \kms at a distance of 15.2 Mpc, while the main 
maser components are blueshifted between 950 - 990 \kmss. Although initially the string of maser 
components has been interpreted as part of a rotating disk, the component velocities and the offset 
arc-structure do not support that picture. 
Instead, it appears that the maser components are a shocked part of the nuclear ISM that is also seen in 
blueshifted OH and HI components \cite{HagiwaraEA2004}. These components are possibly connected 
to the two super-starburst regions found East of the core \cite{MiddelbergEA2007}, which appear 
to be associated with the nuclear blowout seen in the optical. 
Shocks passing through the nuclear ISM provide for the H$_2$O population inversions 
resulting in the amplification of diffuse radio background across the nuclear region \cite{BaanIrwin1995},
which in turn will result in concentrated regions of diffuse and compact emission. 

The cross-correlation spectrum of NGC\,3079 from the RAO-GBT observation at 2.3 ED has been 
presented in Figure \ref{fig1}a and shows two features peaking at 963 and 955 \kmss. 
The strength of the features on the space-Earth baseline is significantly lower than obtained on the
terrestrial baseline between GBT and Effelsberg in Figure \ref{fig1}b.  The other features in the 
terrestrial spectrum were not detected on the space-Earth baseline and no detections have been made 
for NGC\,3079 at any longer baselines.

The decrease in strength of the detected features and the fact that no further detections were made 
at longer baselines would indicate that the maser emission is mostly extended at a 2.3 ED baseline, and 
appears completely resolved at longer baselines. 
The beam size at 2.3 ED suggests that the strongest masering components in the nuclear medium are 
larger than 1400 AU at the distance of NGC\,3079, which is consistent with amplification by diffuse medium.
Although the association with shocks resulting from the super-starburst regions is not confirmed, 
any change in the velocity and spatial structure of the maser components found in past VLBI observations 
may help to confirm the nature of the excitation of these masering regions.

\begin{figure}[h]
\begin{center}
\includegraphics[width=7cm]{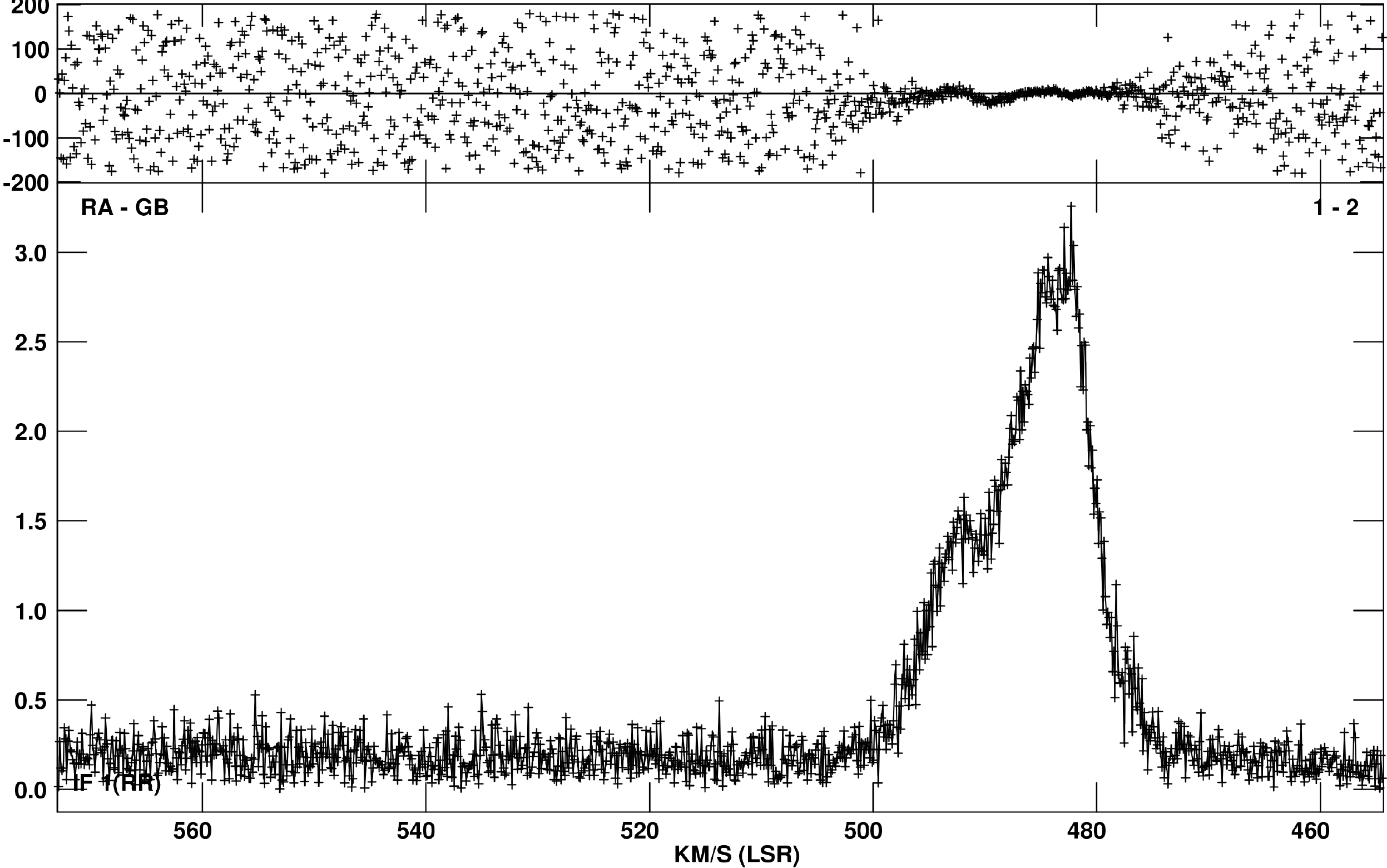} 
 \caption{RadioAstron detection of NGC\,4258. The uncalibrated cross-correlation spectrum of the 
 RAO-GBT baseline of 1.9 ED. Flux density (arbitrary units) is plotted versus radial velocity.}
   \label{fig2}
\end{center}
\end{figure}

\section{NGC\,4258 results}

The H$_2$O MM emission regions in NGC\,4258 are confined to a nearly edge-on disk of 0.5 pc 
surrounding the nuclear AGN  \cite{HerrnsteinEA1998}, also qualified as a CSO.
The orbiting molecular regions within the disk drift in front of the southern part of the 
CSO radio continuum and amplify this continuum. Because of the orbital motion in the disk, 
the masering components drift across the spectrum from low velocity to high, at approximately 8.1 km 
s$^{-1}$ yr$^{-1}$ across the velocity range  440 - 550 \kms \cite{HaschickBP1994,HumphreysEA2008}. 
The systemic velocity of NGC\,4258 is 472 \kms at a distance of  (approximately) 7 Mpc and at half the distance 
to NGC\,3079.

At the time of this writing, the H$_2$O MM emission in NGC\,4258 has been detected with 11 
RadioAstron experiments, the first dating back to 2014. While  fringes were initially
found in observational data at a baseline of 1.9 ED, the updated orbital model of RAO at the ASC correlator 
resulted in subsequent detection of fringes up to baselines of 
26.7 ED (corresponding to 340,000 km). The detection of fringes of the H$_2$O MM emission on this long
RAO-GBT baseline constitutes an absolute record of 8 $\mu$as in angular resolution. 

The RAO-GBT cross correlation spectrum for NGC\,4258 at a baseline of 1.9 ED is presented in Figure 
\ref{fig2}. This (uncalibrated) spectrum shows a two-component profile that resembles the one
obtained with terrestrial baselines, except for the lower amplitude on the space-Earth baseline.  
The resolution obtained for this baseline is 110 $\mu$as, which corresponds to 790 AU at the distance of 
NGC\,4258. 
The profile also shows that a large fraction of the emission regions has not yet been resolved at this resolution, 
which is different from the results obtained for NGC\,3079 at lower spatial resolution.

\begin{figure}[h]
\begin{center}
\includegraphics[width=6cm]{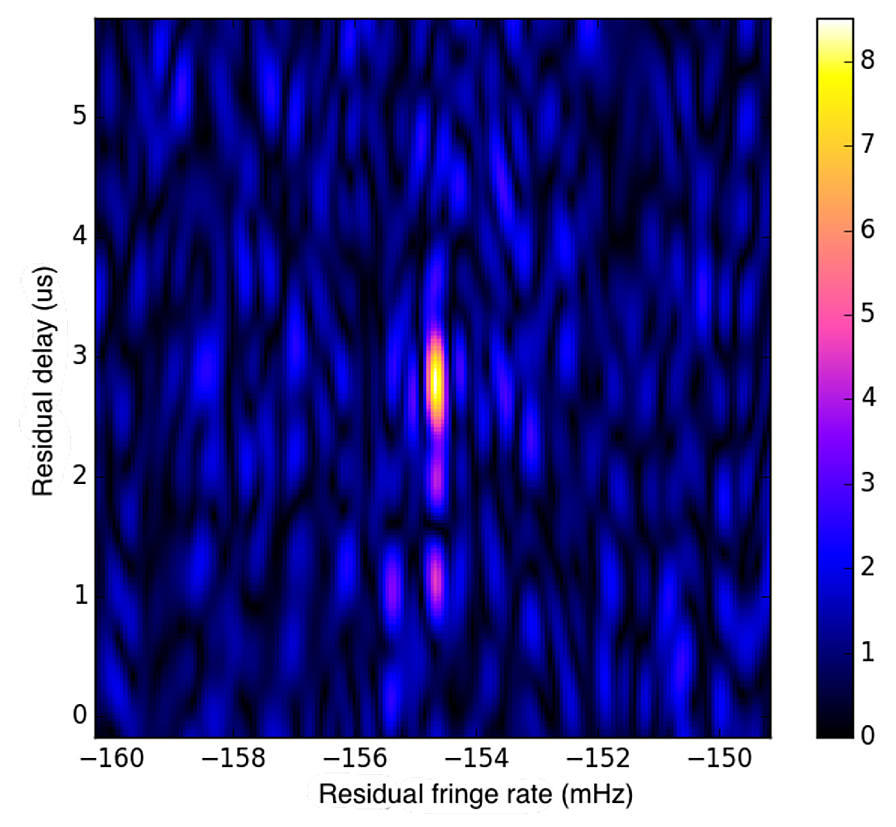} 
 \caption{The fringe amplitude plot of the RAO-Medicina detection of NGC\,4258 at 26.7 ED. 
 The ratio of the interferometer fringe amplitude to the average noise amplitude is plotted 
 against residual delay and fringe rate. }
   \label{fig3}
\end{center}
\end{figure}

At higher resolution an increasing part of the diffuse maser components in NGC\,4258 will be resolved, and only
more compact components will remain unresolved. This is evident in the fringe amplitude plot of the 
detection with the 26.7 ED RAO-Medicina baseline displayed in Figure \ref{fig3}. Three 
individual components may be identified in this plot with a spatial resolution of 56 AU at the 
distance of NGC\,4258.
The mere detection of such compact masering components in NGC\,4258 provides stringent limits on
the degree of saturation and the excitation process. In addition, these more compact masering 
regions are likely to have less tangled magnetic fields and may allow detection of the magnetic field strength 
by its polarization properties.

\section{Overview}

The RadioAstron space-VLBI mission has successfully detected extragalactic H$_2$O MegaMaser 
emission regions, at space-Earth baselines ranging from 1.4 to 26.7 ED.  
The preliminary results for NGC\,3079 and NGC\,4258 at shorter baselines already indicate masering environments and excitation conditions that are distinctly different for the two galaxies. Although NGC\,3079 has not been detected at baselines longer than 2.3 ED, early results for NGC\,4258 suggest that individual masering regions can be detected at longer baselines up to 340,000 km.

\section{Acknowledgements}
The RadioAstron project is led by the Astro Space Center of the Lebedev Physical Institute of the Russian Academy of Sciences and the Lavochkin Scientific and Production Association under a contract with the Russian Federal Space Agency, in collaboration with partner organizations in Russia and other countries.
These results are based partly on observations with the 100-m telescope of the MPIfR (Max-Planck-Institute for Radio Astronomy) at Effelsberg, on observations with the Medicina telescope operated by INAF - Istituto di Radioastronomia, and on observations with the 110-m 
Green Bank Observatory (GBT), which is a facility of the National Science Foundation operated by Associated Universities, Inc., under a cooperative agreement.
Results from the optical positioning measurements of the Spektr-R spacecraft by the global MASTER Robotic 
Net \cite{Lipunov2010}, the ISON collaboration, and the Kourovka observatory were used for spacecraft orbit 
determination.

\end{document}